\magnification=1200
\def\dsp{\baselineskip=12pt plus 1 pt minus 1 pt}

\dsp
\def\title#1{\centerline{\bf #1}}
\def\author#1{\bigskip\centerline{#1}}
\def\address#1{\centerline{#1}}
\def\sec#1{\bigskip\centerline{{\bf #1}}\bigskip}
\def\subsec#1{\medskip\centerline{{\it #1}}\medskip}

\def\cm{{\rm cm}}

\def\K{{\rm K}}

\def\km{\rm km }
\def\yr{\rm yr }
\def\s{{\rm s} }

\def\v#1{{\bf#1}}

\def\cf{{\it cf.~}\/}
\def\eg{{\it e.g.~}\/}
\def\ie{{\it i.e.~}\/}
\def\etal{{\rm et al.~}\/}

\def\sun{\odot}
\def\go{
\mathrel{\raise.3ex\hbox{$>$}\mkern-14mu\lower0.6ex\hbox{$\sim$}}}
\def\lo{
\mathrel{\raise.3ex\hbox{$<$}\mkern-14mu\lower0.6ex\hbox{$\sim$}}}
\bigskip
\title {On the Profiles and the Polarization of}
\title{Raman Scattered Emission Lines in Symbiotic Stars}
\title{II. Numerical Simulations}
\author{K. W. Lee$^1$ \& Hee-Won Lee$^2$}
\author{$^1$Dept. of Astronomy and Atmospheric Sciences}
\address{Kyungpook National University, Taegu, Korea}
\author{$^2$ Research Institute for Basic Sciences and Dept. of Astronomy} 
\address{Seoul National University, Seoul, Korea}
\author{email: hwlee@astro.snu.ac.kr}

\sec{ABSTRACT}

A Monte Carlo method is used to calculate the profiles and the 
polarization of the Raman scattered
O~VI lines ($\lambda\lambda 6827, 7088$) in symbiotic stars, which
are believed to be a binary system of a cool giant and a hot star
with an emission nebula around it.
A point-like isotropic UV radiation source is assumed and
a simple spherical wind model is adopted for the kinematics of the scattering 
material from the cool giant.

We first investigate the case where the incident line photons are described by a
Gaussian profile having a width of $10^4~\K$. We subsequently investigate
the effects of the extended ionized region and non-spherical wind models 
including a disk-type wind and a bipolar wind. The cases where the 
emission source is described by
non-Gaussian profiles are briefly studied. 

Finally as an additional
component for the kinematics of symbiotic stars
the orbital motion of the hot component around the cool giant is included 
and the effect on the spectropolarimetry is investigated.
In this case the polarization direction changes around the red
part of the Raman-scattered emission lines, when the observer's
line of sight is perpendicular to the orbital plane, and no such
effect is seen when the line of sight lies in the orbital plane. 
Furthermore, complex peak structures are seen in
the degree of polarization and the polarized flux,
which have often been observed in several symbiotic systems including
RR Tel.

Brief observational consequences and predictions
are discussed in relation to the present and future
spectropolarimetry for symbiotic stars.  It is concluded that
spectropolarimetry may provide a powerful diagnostic of the
physical conditions of symbiotic stars.

\sec{1. Introduction}

A symbiotic star is believed to be a binary system consisting of a cool
giant and a hot star providing H-ionizing photons for an emission nebula
around it (Iben \& Tutukov 1996). It is expected that the hot component is
characterized by a typical temperature $T\sim 10^{4-5}~\K$ and that the
binary motion has a typical period of $P\sim 10^{2-3}~{\rm days}$.

The observed spectral parts of the emission in the symbiotic systems cover
a very large range including radio, IR, optical, UV and X-rays.
The complicated profiles and the significant variabilities
often seen in the UV emission lines imply that
the dynamics and the physical conditions are not simple enough
to be described by a unique model (\eg Vogel \& Nussbaumer 1994).

It has been known that a significant fraction of symbiotic stars exhibit
broad emission features around $\lambda 6827$ and $\lambda 7088$ with
widths up to $20~{\rm\AA}$, which are an order of magnitude larger than
those of other emission lines in their typical spectra (\eg Allen 1980). 
These broad features 
are identified by H. Schmid (1989) as the Raman scattered O~VI $\lambda
\lambda 1032,~ 1038$ by hydrogen atoms. During the scattering process,
the atom in the ground state absorbs the incident line photon 
with frequency $\nu_i$ to be excited to an
intermediate state followed by de-excitation to the 2$s$ state emitting a
photon with frequency $\nu_f=\nu_i-\nu_\alpha$, where $\nu_\alpha$ is the
frequency of the Ly $\alpha$ transition.

The Raman scattering nature
explains the large width of the features and many observational
characteristics of symbiotic stars such as the existence of highly
ionized lines. Observational confirmations of the Raman scattering mechanism
include the spectropolarimetry which shows
that the features are highly polarized in contrast with other negligibly
polarized lines (\eg Schmid \& Schild 1994). The identification of
the Raman scattering nature is corroborated by the UV observation
by Espey \etal (1995), who showed that the symbiotic star RR Tel exhibits
a very strong O~VI doublet with the intense broad features
around $\lambda 6827$ and $\lambda 7088$.

Valuable kinematic information on the emission nebula of a symbiotic system
is obtained in a detailed analysis of the profiles of the emission lines,
which often accompany non-trivial structures including double-peaked or partial
absorption troughs (\eg Mueller \& Nussbaumer 1985, Pereira \etal 1995). 
The Raman scattered features can provide a useful diagnostic to constrain many
kinematical parameters of a symbiotic system,
because they have broadened profiles due to the relative motion
of the stellar wind around the cool giant with respect to the emission
nebula. Furthermore, the width of the profiles is also enhanced by the 
inelastic nature of the scattering by almost an order of magnitude.

Recently Harries \& Howarth (1996, 1997) provided a good amount of
spectropolarimetric data and performed a fairly complete Monte Carlo
calculation about the Raman scattered emission features (see also
Schmid 1996). In their observational data some symbiotic stars show
the polarization direction flip around the red part of the Raman scattered
feature. This kind of behavior usually accompanies multiple peak 
structures in the polarized flux (\eg Espey \etal 1995, Schmid 1996). However, 
in the spherical wind model, many Monte Carlo computations show that 
the polarization flip usually occurs around the center of the feature 
calling a need to investigate non-spherical models.

This is the second paper in a series on the polarization and the line
profiles of the Raman scattered flux.
In Lee \& Lee 1997 (hereafter Paper I), we reviewed the basic atomic
physics of the Raman scattering 
of incident photons shortward of Ly~$\alpha$ by hydrogen atoms, 
and computed the profiles and the polarization for a few cases 
using the single scattering approximation (see also Lee \& Lee 1996).
A more complete review
about the atomic physics is provided by Isliker \etal (1989) and
the references therein.
In this paper we calculate the line profiles and the polarization of the
Raman scattered O~VI lines for various kinematic models using a Monte 
Carlo method.

The paper is composed as follows. In section 2, the basic procedures of the
Monte Carlo method and the model descriptions are given. Using the Monte Carlo
code we investigate some fundamental properties of the Rayleigh-Raman scattering
processes
and discuss them in section 3. The results of the numerical calculations
are presented in section 4. In section 5, observational implications are 
commented. Finally, in section 6, a summary is presented with discussions about
the significance of spectropolarimetry in understanding the symbiotic systems.

\sec{2. Model}

There has been much observational work on symbiotic systems, showing that
complicated kinematics is responsible for variabilities and complex structures
in the line profiles (\eg Mueller \& Nussbaumer 1985, Pereira \etal 1995). 
We do not attempt to give a detailed dynamical model
in this work, but adopt rather simple models in order to find out
the main features contributing to the polarization structures and the profiles
in the Raman scattered lines (\eg Harries \& Howarth 1996, Schmid 1996). 
In particular, the polarization behavior
is expected to depend upon the scattering optical depth of the system,
which naturally measures the number of scatterings before a given
photon gets into the line of sight. The Raman scattering
of the O~VI photons is also characterized by an enhanced Doppler effect,
according to which the wavelength shift of an incident photon increases
by a factor of $\sim$ 7. Therefore, the kinematics of the scatterers 
may combine with the scattering optical depth structure to contribute
to the complicated behavior of the polarization of the Raman scattered
features.

In this section, we summarize the basic points of the Monte Carlo code
and the kinematic models adopted in our numerical simulations.
For the Raman and Rayleigh scattering cross sections of the O~VI doublets,
we use the results in Paper I, in which we showed that
$$\eqalign{
\sigma_{Ram}(1032)=7.5\ \sigma_T ~,~ \sigma_{Ram}(1038)=2.5\ \sigma_T \cr
\sigma_{Ray}(1032)=34\ \sigma_T ~,~ \sigma_{Ray}(1038)=6.8\ \sigma_T  \cr
}\eqno(2.1)$$
where $\sigma_T=6.6\times 10^{-25}~\cm^2$ is the Thomson scattering 
cross section.

\subsec{2.1 Monte Carlo Method}

In this subsection we briefly describe the Monte Carlo method to compute
the profiles and the polarization of the Raman scattered O~VI doublet
lines. It is first assumed that there is an emission region embedded in 
the scattering medium consisting of the extended atmosphere of the
cool giant. The incident UV line photons are further assumed to
be emitted isotropically and unpolarized. We start with the incident photons 
governed by a Gaussian profile corresponding to a Maxwell-Boltzmann
distribution. We subsequently investigate the
cases where the incident radiation field is described by synthetic
profiles such as double-peaked profiles often seen in the observational
data.

In Fig.~1 we show a schematic geometry of a symbiotic star system,
which is adopted for our Monte Carlo calculation. 
The coordinate system is chosen in such a way that the binary axis
coincides with the $z$-axis and the observer's line of sight is
the $x$-axis. The hot star
is represented by the circle on the right side, and the bigger
circle in the origin is the cool giant. A spherically symmetric
stellar wind is depicted by radial arrows around the cool giant.

For a numerical simulation of the Rayleigh-Raman scattering it is
essential to compute the free path length at a position for a given
wave vector and wavelength. The mean free path of a UV line photon
is given
$$l_f=(n\sigma_{tot})^{-1}, \eqno(2.2)$$
where the total scattering cross section $\sigma_{tot}$ is defined by
$$\sigma_{tot}\equiv \sigma_{Ram}+\sigma_{Ray}.\eqno(2.3)$$
A given photon travels up to a distance $l$ with a probability
$$p_l=1-\exp(-l/l_f)=1-\exp(-\tau_l), \eqno(2.4)$$
where 
$$\tau_l\equiv l/l_f=\int ds\ n(s)\sigma_{tot}.\eqno(2.5)$$ 
In terms of
a random uniform variable $r$ in an interval between 0 and 1, $\tau_l$
can be generated by a transformation
$$\tau_l =-\ln(1-r), \eqno(2.6)$$ 
identifying $r$ with $p_l$. The inverse transformation of Eq.~(2.5)
gives the free path length of the incident photon.

A photon travels a free path $l$ before a hydrogen atom Raman-scatters
it with a branching ratio $r_{Ram}\equiv \sigma_{Ram}/\sigma_{tot}$.
If a random number is greater than $r_{Ram}$, then we regard this scattering
as Rayleigh, and otherwise the photon is Raman-scattered.

When the scattering type is Rayleigh, the escape condition should be 
examined before the next scattering site is determined.
The escape condition is simply
$$\tau_l \ge \tau_{esc} , \eqno(2.7)$$
where $\tau_{esc}$ is the scattering optical depth to an observer at 
infinity, \ie, 
$$\tau_{esc} =\int_s^{\infty}\  ds\ n(s) \sigma_{tot}.
\eqno(2.8)$$ 
If the escape condition is met, then the UV photon is regarded as being emitted
to infinity. Otherwise a new wave vector and the polarization components are
calculated to be used for the subsequent scattering. 

On the other hand, if the photon is Raman-scattered, then we record the photon 
with the polarization according to the wavelength. Here, we assume that
the continuum absorption around H$\alpha$ is negligible and that
the Raman-scattered
photon escapes freely. The effect of continuum opacity has been 
extensively investigated by Schmid (1992, 1996) and Harries \& Howarth (1997).
In order to keep the standard deviation no larger
than 1.5 percent, we generate typically $5\times 10^6$ photons for each run.

In the following section, we discuss the stellar wind models and calculate
the relevant scattering optical depths and the escape conditions. 

\subsec{2.2  Stellar Wind Models}

We discussed the kinematic models in Paper I, which we adopt and
again introduce in this paper (see also Nussbaumer \& Vogel 1987). A 
typical wind velocity law can be written as
$$\eqalign{
\v v(\v r)&=v_\infty\left(1-{R_*/r}\right)^\beta\hat\v r \cr
&=v_\infty (1-\rho^{-1})^{\beta}\hat\v r ,\cr
}  \eqno(2.9) $$
where $v_\infty$ is the terminal wind velocity, $R_*$ is
the radius of the cool giant, $\rho\equiv r/R_*$ is the radial distance
in the unit of $R_*$, and $\beta$ is a positive constant. 
The density profile corresponding to the
velocity law Eq.~(2.9) is given by 
$$n(\v r) = n_0 \rho^{-2}(1-\rho^{-1})^{-\beta}, \eqno(2.10)$$
where $n_0$ is a typical density given in terms of the mass 
loss rate $\dot M$, and the proton mass $m_p$ by
$$n_0\equiv {\dot M/ 4\pi R_*^2 m_p v_\infty}.\eqno(2.11)$$
We again choose $\beta=1$ as in Paper I. 

Therefore, the total scattering
optical depth between position 1 and position 2 is given by
$$\eqalign{
\tau_{12} &=\int_{s_1}^{s_2} ds\ n(\v r)\ \sigma_{tot} \cr
 &=\tau_{0} \int_{\rho_1}^{\rho_2} {d\rho\over\sqrt{\rho^2-b^2}} 
{1 \over (\rho-1)}, \cr
}\eqno(2.12)$$
where $s$ is the running
parameter along the photon path, and the representative scattering
optical depth
$$\tau_0\equiv n_0 R_* \sigma_{tot}. \eqno(2.13)$$
Here, $b$ is the impact parameter
of the photon path with respect to the center of the cool giant
divided by $R_*$ (see Fig.~1 for a schematic geometry). In this paper, 
it is understood that all the distances
are measured in units of the radius of the cool giant $R_*$.

If the $i$-th scattering site is given by $\vec \rho_i$ and 
the wave vector by $\hat \v k_i$, the next scattering site $\vec\rho_{i+1}$ 
characterized by the scattering optical depth $\tau_i$ from $\vec
\rho_i$ is obviously given by
$$\eqalign{
\vec \rho_{i+1}&=\vec\rho_i+|\vec\rho_{i+1}-\vec\rho_i|\hat\v k_i \cr
&=\vec\rho_i+s_i(\tau_i)\hat\v k_i, \cr
} \eqno(2.14)$$
where $s_i(\tau_i)$ is the distance between the
two scattering sites, which is dependent on the vectors
$\vec\rho_i$, $\hat\v k_i$ and $\tau_i$. 

The functional form of $s_i$
can be written separately for the case where the impact
parameter of the photon trajectory is larger than the radius of the cool 
giant and for the case where it is smaller than the radius of the giant,
\ie, either $b>1$ or $b<1$.

In order to express the functional form of $s_i(\tau_i)$, we
perform an integration of \break Eq.~(2.12). Introducing a function 
$f(\rho)$ defined by
$$f(\rho)\equiv {1 \over \sqrt{|1-b^2|}}\left[\rho + \sqrt{\rho^2-b^2}-1
\right]\eqno(2.15)$$
we have an equivalent relation
$$\rho={1 \over 2} \left[1+\sqrt{|1-b^2|} f(\rho)+b^2\{ 
(1+\sqrt{|1-b^2|} f(\rho)\}^{-1} \right],\eqno(2.16)$$
which gives $\rho$ after $f(\rho)$ is obtained.

If $b<1$, then we obtain
$$\cases{
f(\rho_{i+1})=\displaystyle \coth\left[ \coth^{-1}
f(\rho_i)-{\sqrt{1-b^2} \over 2}{\tau_i\over\tau_0}\right] \cr
\noalign{\vskip6pt}
s_i(\tau_i)=\sqrt{\rho^2_{i+1}-b^2}-\sqrt{\rho^2_i-b^2} \cr}\eqno(2.17)$$
for $\hat\v k_i\cdot\hat\v r >0$, and
$$\cases{
f(\rho_{i+1})=\displaystyle \coth\left[ \coth^{-1}
f(\rho_i)+{\sqrt{1-b^2} \over 2}{\tau_i\over\tau_0}\right] \cr
\noalign{\vskip6pt}
s_i(\tau_i)=\sqrt{\rho^2_i-b^2}-\sqrt{\rho^2_{i+1}-b^2} \cr}\eqno(2.18)$$
for $\hat\v k_i\cdot\hat\v r <0$.

On the other hand, if $b>1$, then the result is
$$\cases{
f(\rho_{i+1})=\displaystyle \tan\left[ \tan^{-1}
f(\rho_i)+{\sqrt{b^2-1} \over 2}{\tau_i\over\tau_0}\right] \cr
\noalign{\vskip6pt}
s_i(\tau_i)=\sqrt{\rho^2_{i+1}-b^2}-\sqrt{\rho^2_i-b^2} \cr}\eqno(2.19)$$
for $\hat\v k_i\cdot\hat\v r >0$.
If $b>1$ and $\hat\v k_i\cdot\hat\v r <0$, 
then we introduce the scattering optical depth $\tau_b$ to the
impact point, which is given by
$$\tau_b={2 \tau_0 \over \sqrt{b^2-1}}\left[ \tan^{-1} f(\rho_i)
-\tan^{-1} f(b)\right]. \eqno(2.20)$$
Then the final result is written as
$$\cases{
f(\rho_{i+1})=\displaystyle \tan\left[ \tan^{-1}
f(\rho_i)-{\sqrt{b^2-1} \over 2}{\tau_i\over\tau_0}\right] \cr
\noalign{\vskip6pt}
s_i(\tau_i)=\sqrt{\rho^2_i-b^2}-\sqrt{\rho^2_{i+1}-b^2} \cr}\eqno(2.21)$$
for $\tau_i < \tau_b$, and
$$\cases{
f(\rho_{i+1})=\displaystyle \tan\left[ \tan^{-1}
f(b)+{\sqrt{b^2-1} \over 2}{\tau_i\over\tau_0}\right] \cr
\noalign{\vskip6pt}
s_i(\tau_i)=\sqrt{\rho^2_i-b^2}+\sqrt{\rho^2_{i+1}-b^2} \cr}\eqno(2.22)$$
for $\tau_i > \tau_b$.

The escape conditions are met when the scattering optical depth
$\tau_i \ge \tau_{esc}$. Here, the escape scattering optical depth $\tau_{esc}$
is obtained from Eqs.~(2.17)-(2.22) by letting $s_i \to \infty$, \ie, 
$$\tau_{esc}=\cases{\displaystyle{2 \tau_0 \over \sqrt{1-b^2}}
\coth^{-1} f(\rho_i), &for $b<1,~ \hat\v k_i\cdot\hat\v r >0$; \cr
\noalign{\vskip6pt} \displaystyle
{2 \tau_0 \over \sqrt{b^2-1}}\left[ {\pi \over 2}-\tan^{-1}
f(\rho_i)\right], & for $b>1,~ \hat\v k_i\cdot\hat\v r >0$; \cr
\noalign{\vskip6pt} \displaystyle
{2 \tau_0 \over \sqrt{b^2-1}}\left[ {\pi \over 2}-\tan^{-1}
f(b)\right], & for $b>1,~ \hat\v k_i\cdot\hat\v r <0$. \cr}\eqno(2.23)$$
No escape is possible for the case where $b<1$ and
$\hat\v k_i\cdot\hat\v r <0$, because the total optical depth
to the surface of the cool giant diverges for all $\rho_i$. This means that
the photons are scattered off before they reach the surface of the cool giant.

The final procedure of the Monte Carlo simulation is to determine the
Doppler factors and the polarization associated with the scattered photon. 
The polarization state of an ensemble of photons is described by a
density matrix (\eg Berestetskii \etal 1971, Lee \etal 1994), where
the Stokes parameters $Q$, $U$, and $V$ correspond to the difference
of the main diagonal elements, the real and the imaginary parts of
the off-diagonal elements of the density matrix, respectively. 
The Rayleigh-Raman scattering is characterized by the Rayleigh phase function,
and the wave vector and the polarization of the scattered photon are 
computed by the same way as in the case of the Thomson scattering.

For the Raman scattering of the O~VI doublets,
the total wavelength shift is given by
$${\Delta\lambda_f\over\lambda_f}={\lambda_f\over\lambda_i}
{{\hat\v k_i\cdot\v v(\v r)}\over c}+
{{\hat\v k_f\cdot\v v(\v r)}\over c} \eqno(2.24)$$
where $\hat\v k_i$ is the wave vector of the incident photon and
$\hat\v k_f$ is that of the outgoing photon. Eq.~(2.24)
is also valid for Rayleigh scattering, where $\lambda_f=\lambda_i$. It is
clear that in the case of Raman scattering the Doppler factor becomes
large by a factor of about 7 due to the motion of the scatterers relative to 
the emission source. We show the contours of constant total scattering optical 
depth and Doppler factors in Fig.~2, which were introduced and discussed
in Paper I.

\sec{3. Fundamental Properties of Raman Scattering}

In this section we discuss the characteristic features of the Raman scattering
process using the Monte Carlo method. Schmid (1992, 1995, 1996) also discussed
the basic properties of the Rayleigh-Raman scattering process 
(see also Harries \& Howarth 1997) and we elaborate further the 
properties of the scattering process in a similar way to deal with 
the random walk process.

\subsec{3.1 Single Scattering Approximation}

In Paper I, we briefly mentioned the single scattering approximation, which 
applies to the case where the scattering region is characterized by a
small scattering optical thickness. This formalism can also be
useful as a check of our Monte Carlo code, in the sense that it is
the limiting case when the total scattering optical depth tends to zero.

In Fig.~3 we present a result from our Monte Carlo code for
a scattering optical depth $\tau_0=0.5$ and compare it with the corresponding
result from the single scattering approximation. The 
adopted parameters are described in Paper I.
The Monte Carlo result is in good agreement with the result
from the single scattering approximation within one standard deviation
shown by the error bars in the degree of polarization. 

In the following subsection we quantify the single scattering approximation
and investigate some fundamental properties of the Rayleigh-Raman scattering process.

\subsec{3.2 Reflection from a Slab}

A UV photon with a wavelength shorter than 1216 \AA~ can be scattered either
by a Rayleigh process or by a Raman process from a hydrogen atom. Therefore,
the UV photon may be Rayleigh-scattered several times before it is 
Raman-scattered. Under the assumption that a Raman-scattered photon has
a very small optical depth, it will escape freely from the scattering region
and may reach the observer.
The ratio of the number of the emergent photons which are Raman-scattered 
to that of the total incident photons is expected to be dependent 
on the ratio of the Raman scattering cross section to the total scattering 
cross section and the scattering geometry such as the total optical depth of
the medium. 

In this subsection, we discuss briefly some of the fundamental properties
of the Rayleigh-Raman scattering process of UV line photons in a simple 
scattering medium.

H. Schmid, who proposed the Raman process in symbiotic systems and performed 
pioneering works in this subject, presented the basic results in his papers 
(\eg Schmid 1992, Schmid 1996).
A Monte Carlo simulation is particularly useful in describing the scattering
process, because we can conceptually divide the emergent photons according
to their scattering numbers and perform a detailed analysis. Pursuing in 
this line of reasoning, we collect the photons reflected from the both sides
of a slab of finite scattering optical depth, using the Monte Carlo code.
It is assumed that the slab is illuminated from outside and that the distance
from the slab to the incident raditation source is much larger than the size
of the slab, so that the incident photons enter the slab effectively normally. 
The emergent photons
are subsequently divided into their number of scatterings,
and we record their flux and the degree of polarization for
further analysis.

Fig.~4 illustrates the result from the Monte Carlo calculation on
the Rayleigh and
Raman reflected components from a finite slab as a function of the number of
scatterings. On the vertical axis is shown the logarithm of the number of the
photons reflected from the slab. 
The total scattering optical depth  $\tau_s$ of the slab is 
chosen to be $\tau_s=0.5,~1,~5,~10.$
The ratio $r_{Ray}\equiv \sigma_{Ray}/\sigma_{tot}$ of the Rayleigh scattering 
cross section to the total 
scattering cross section is taken to be $r_{Ray}=0.2,~0.8$, where
the case $r_{Ray}=0.2$ is represented by the light lines and the thick lines
are for the $r_{Ray}=0.8$ case. 

When the scattering optical depth is small and $r_{Ray}=0.8$, the 
Rayleigh-scattered
flux is larger than the Raman-scattered flux, whereas the converse
is true for large scattering optical depths. This is because 
the Rayleigh-scattered photons are trapped 
as the scattering optical depth increases, whereas
all the Raman-scattered photons are assumed to escape on the spot.

As is usually expected the main contribution to the scattered flux is
due to singly scattered photons. Both the flux and the degree of polarization
show exponential decrease as a function of the number of scatterings. This
behavior is the foundation of the single scattering approximation discussed
in the previous section and in Paper I.

We may give a semi-quantitative argument about the fraction of the Raman 
scattered flux to the total incident flux.
For a given slab of total scattering optical depth $\tau_s$, a fraction
$(1-e^{-\tau_s})$ will be scattered at least once. Let's denote the total
incident number flux by $f_0$. Then $f_s\equiv f_0(1-e^{-\tau_s})$ is the total
scattered (number) flux. Let $f(n)$ be the number flux scattered no less than
$n$. From this definition, it is obvious that
$$f(1)=f_s. \eqno(3.1)$$

Furthermore, a fraction $(1-r_{Ray})$ of $f_s$ will be Raman scattered and
escape the region. Hence if we denote by $f_{Ram}(n)$ the Raman scattered
(number) flux scattered only $n$ times, then
$$f_{Ram}(1)=(1-r_{Ray})f_s=(1-r_{Ray})f(1). \eqno(3.2)$$
In fact, this relation holds for any $n$, that is,
$$f_{Ram}(n)=(1-r_{Ray})f(n), \eqno(3.3)$$
because any Raman scattered photon does not suffer a subsequent scattering.

In a similar way we denote by $f_{Ray}(n)$ the Rayleigh scattered emergent
number flux scattered only $n$ times. Then we may write
$$f_{Ray}(n)=r_{Ray}\beta(n) f(n). \eqno(3.4)$$
Here, the function $\beta(n)$ is loosely defined as an escape probability
from the $n-$th scattering site. Because the scattering process is
similar to the random walk process, there is no definite site of $n-$th
scattering, and we can only state in a probabilistic way. If we borrow
from the solution of the random walk problem, we may state that the photon
diffuses up to the position where the scattering optical depth from the
surface is approximately $\sqrt{n}$. Therefore, we may tentatively set 
$$\beta(n) = {1\over 2}(e^{-\sqrt{n}}+e^{-(\tau_s-\sqrt{n})}).\eqno(3.5)$$
When the number of scattering exceeds $\tau_s^2$, then the above formula
does not give any meaningful results. However, the flux $f(n)$ decreases
approximately exponentially, the result will not be sensitive to the choice
of $\beta(n)$ for sufficiently large $n$.

Having defined $f_{Ram}(n),$ $f_{Ray}(n),$ and $\beta(n)$, we can formally
give a recursion formula for $f(n)$, which is
$$\eqalign{
f(n+1) &=f(n)-f_{Ram}(n)-f_{Ray}(n) \cr
       &=r_{Ray}[1-\beta(n)]f(n). \cr
}\eqno(3.6) $$
The functional dependence of $\beta(n)$ on the scattering optical depth and
the scattering geometry determines the basic properties of the Rayleigh-Raman
scattering process.

In Fig.~5 we plot $f_{Ram}(n),~f_{Ray}(n)$ as a function of the
scattering number $n$, which are obtained recursively using Eq.~(3.6).
A comparison is made with the corresponding Monte Carlo results. 
The solid lines show the Monte Carlo results and the dotted lines
give $f_{Ram}(n)$ and $f_{Ray}(n)$ obtained analytically. 
Eq.~(3.5) is used for the escape probability for the upper panel.  
In the bottom panel we show a different set of $f_{Ram}(n)$ and $f_{Ray}(n)$ 
using another formula for the escape probability, defined by
$$\beta'(n)={1\over 2} \exp(-n^{0.4}). \eqno(3.7)$$
The adopted
parameters are $\tau_s=10$, and $r_{Ray}=0.8$, which is relevant for the
O~VI doublets $\lambda \lambda 1032, 1038$. Due to the choice of the parameters
we see that the Raman-scattered fluxes are larger than the Rayleigh-scattered
counterpart. As is shown in Fig.~5a, 
the agreement is very good except when the number of scatterings becomes large,
where the small remainder flux does not give much significance. 
Far better agreement is obtained in Fig.~5b. Considering the agreement 
shown in Fig.~5, the preceding argument may be said to closely 
describe the scattering process in the scattering medium.

The total Raman scattered flux $F_{Ram}$ is given by
$$F_{Ram}=\sum_n f_{Ram}(n)=(1-r_{Ray})\sum_n f(n). \eqno(3.8)$$
Similarly, we may write the total Rayleigh scattered flux $F_{Ray}$ 
$$F_{Ray}=\sum_n f_{Ray}(n)=r_{Ray}\sum_n \beta(n) f(n). \eqno(3.9)$$
Hence, we can estimate the ratio $R_{Ram}$ of the total Raman scattered 
flux to that of the Rayleigh scattered flux 
$$\eqalign{
R_{Ram}&\equiv F_{Ram}/F_{Ray} \cr
       &= {(1-r_{Ray})\over r_{Ray}} R_s, \cr
} \eqno(3.10)$$
where 
$$R_s={\sum_n f(n) \over \sum_n \beta(n) f(n)}. \eqno(3.11)$$

For sufficiently large $\tau_s$, $\beta(n)$ decreases for small $n$. Because
$f(n)$ is decreasing sufficiently fast, the ratio $R_s$ is
dominantly affected by the sum of the first few terms. A conservative
estimate can be obtained by considering only the first term, which gives
$$R_{Ram}^{(1)} = {(1-r_{Ray})\over \beta(1)\ r_{Ray}}. \eqno(3.12)$$
Here, in the limiting case where $\tau_s\rightarrow\infty$,
$\beta(1)$ can be evaluated exactly from the first principle, which is simply
$$\eqalign{
\beta^{exact}(1)&=\int_0^\infty dt~e^{-t}\int_0^1 d\mu~{3\over 8}(1+\mu^2)
e^{-t/\mu} \cr
&= (11-12 \ln 2 )/16 \sim 0.168. \cr}
\eqno(3.13)$$
This is approximately equal to $(2e)^{-1}$ given either by Eq.~(3.5) or by Eq.~(3.7).
Hence for $r_{Ray}=0.8$, which
is close to the true value of the O~VI doublets, we have 
$R_{Ram}^{(1)}=1.48$ using $\beta^{exact}$. This ratio is a typical one
when the scattering optical depth is moderately larger than 1.

For a slab of $\tau_s=10$, if we include terms up to $n=20$, 
the ratio becomes $R_{Ram}=2.11$ which differs with $R_{Ram} ^{(1)}$ significantly. 
Note that this number is sensitively
dependent on the total scattering optical depth $\tau_s$ of the slab and
that it usually increases with $\tau_s$.
This implies that even though the Rayleigh scattering cross section is much
larger than that of the Raman scattering cross section, the emergent Raman
scattered flux is comparable to or larger than that of the emergent
Rayleigh scattered flux, when the medium is not optically thin. 
This is also confirmed by Schmid (1992)
and by Harries \& Howarth (1997).

For a small scattering optical thickness, the flux ratio of the reflected 
components is fairly close to the ratio of the scattering cross sections
because the effect of multiple scattering is negligible. 
In the above formalism, this can be shown in a straightforward way. That is,
$f(n)\sim 0$ for $n \ge 2$, and we may take $\beta(1)\sim 1$. Therefore, we have
$$R_{Ram} \sim R_{Ram}^{(1)}={1-r_{Ray}\over r_{Ray}}=
{\sigma_{Ram}\over\sigma_{Ray}}, \eqno(3.14)$$
which implies that no conversion to Raman scattering from Rayleigh scattering occurs
significantly.

\sec{4. Numerical Results}

In this section, we discuss the results of our Monte Carlo calculation 
for various cases including non-spherical wind models. The largest polarization
is expected for $90$ degree scattering, and negligible polarization is obtained
for forward and backward scatterings. In this section the photons collected 
from the numerical simulations are those escaping perpendicular to the binary
axis unless stated otherwise. Because of the symmetry about the binary axis in
spherical wind models the polarization direction is either along the binary
axis or perpendicular to it. Therefore we denote by negative $P$ the polarization along
the binary axis and by positive $P$ the polarization perpendicular to it.

\subsec{4.1 Simple Gaussian Case}

The physical conditions of the emission line region are not known in
detail, and we assume that
it is in thermal equilibrium with a
temperature $T_{em}$, being located very near to the hot star.
Under this assumption, the profile of the
emission line is described by a Gaussian with the width of
$$v_{th}=\sqrt{2kT_{em}/m_a}=2.3\ T_{em,4}^{1/2} ~\km~\s^{-1},\eqno(4.1) $$
where $T_{em,4}\equiv T_{em}/(10^4~\K)$, and the oxygen atomic mass $m_a=16\ m_p$,
$m_p$ being the proton mass. We take the wind terminal velocity
$v_{\infty}=20~\km~\s^{-1}$, $R_*=50 R_\sun$ and the separation of the
cool giant and the hot star $a=10 R_*$ as in Paper I. From now on, 
we will refer this case as the simple Gaussian case, which will serve 
as a reference model throughout the paper.

In Fig.~6  we show the line profiles and the polarization for both 
$\lambda 1032$ and $\lambda 1038$ photons for the cases where 
$\tau_0=0.5, 1.0,10$.

The polarization behavior and the line profiles are similar to those of the single
scattering case, where the large polarization perpendicular to the
binary axis is shown with small flux in the blue part of the feature and 
the polarization becomes small in the center and red parts of the feature.
Furthermore, the polarization direction changes around the center 
part of the feature when $\tau_0\go 5$. 
When $\tau_0\lo 1$, the total flux shows a double-peaked profile
and in the case of opposite limit, a broad single-peaked profile is obtained.
This is already summarized in Paper I and also by Harries \& Howarth 
(1997) (see also Schmid 1996). 

It is noted that the single scattering approximation
gives a good qualitative description when
$\tau_0\lo 10$. Another important point to note
is that the scattering region corresponding to the polarization flip
is the region which is characterized by the Doppler factor about zero,
that is, the sphere having the binary axis connecting the hot component
and the cool giant as a diameter. We may give a simple intuitive explanation 
for this phenomenon in the context of the single scattering approximation as 
follows.

The boundary region for the polarization direction flip 
is the conical region with 
the opening angle of $\pi/4$ with the apex at the hot component. 
From Fig.~2, the scattering region responsible for the blue shift is mostly
inside the conical region giving strong polarization perpendicular to the
binary axis. However, in the center-red part there is a competition
of the two polarization components because the region with Doppler 
factors 0 and larger lies both inside and outside
the conical region. The contribution outside the conical region increases
as the typical scattering optical depth $\tau_0$ becomes large. This
explains the phenomenon that the polarization flip around the central region
is obtained clearly when $\tau_0$ is significantly larger than 1.

\subsec{4.2 Extended Ionized Region}

The assumption that the UV emission line region is spatially point-like is
rather naive and the scattering region reduces severely due to the extended
emission line region.  Several researchers investigated the detailed shape
of the emission nebula around the hot component of a symbiotic
system. Seaquist \etal (1984) provided the shape of the ionization
front using a parameter $X_H$ defined by
$$
X_H\equiv {4\pi a L_H \over \alpha_B}\left({m_p v \over \dot M}^2\right)
\eqno(4.2)$$
where $L_H$ is the total luminosity of the ionizing radiation,
$\dot M$ is the mass loss rate of the cool giant, $v$ is the 
wind speed, $a$ is the separation of the cool giant and the hot component, 
and $\alpha_B$ is the case B recombination coefficient for hydrogen. 
Monte Carlo calculations for various values of $X_H$ have been done 
for the Rayleigh scattering of UV radiation (Schmid 1995) and 
also for the Raman scattering of the O~VI lines (Schmid 1996, Harries \& 
Howarth 1997). We show some of our 
results to point out the profile and polarization features affected by 
the parameter $X_H$. 

We consider three cases where the parameter $X_H$ is given by $X_H=0.04,~
 0.4,~ 4.0$.  For a value of $L_H=130 L_\sun$, where $L_\sun$ is the
solar luminosity, the corresponding mass loss rates of the cool giant
are $\dot M=10^{-7},~10^{-6},~10^{-5}~M_\sun~\yr^{-1}$.
In Fig.~7 is shown a typical ionization structure around the hot star.
The solid lines represent the ionization front for the H~II region, and the
dashed lines are the conic sections which we adopt to approximate the 
true location of the front (\ie the solid lines) in order 
to speed up the numerical calculation.

The O~VI line photons originate deep inside the H~II region, because
O~VI has much higher ionization potential than H~I does. Therefore, 
we assume that the
O~VI region is well-localized around the hot star, having a small size 
compared with that of a typical scattering geometry and justifying the 
point-source assumption.
The results are presented in Fig.~8, where we only show the cases
$\tau_0=1,10$ for the $\lambda 1032$ photons.

The overall results are qualitatively similar to those of the simple Gaussian
case described in the preceding section. However, when $X_H$ gets as large 
as $\sim 4$,
the polarization is dominated by the component perpendicular to the
binary axis, and no polarization flip is seen around the center
of the feature. 
A simple explanation of this behavior is that the extended H~II region
has more intersection with the scattering region responsible for the
parallel component of the polarization than the region giving the perpendicular
component in the simple Gaussian case (\cf section 4.1). Therefore,
compared with the simple Gaussian case, the reduced scattering region for
the parallel component results in the perpendicular polarization on the
whole. This tendency is intensified as $X_H$ increases.
Hence combined with the analysis of section 4.1 the polarization perpendicular
to the binary axis is stronger as $X_H$ increases and $\tau_0$ decreases,
which is displayed by the lower left panel of Fig.~7.

\subsec{4.3 Disk Wind}

As a non-spherical model a disk wind is invoked and described in this 
subsection. For ease of Monte Carlo coding, we just exclude the conic
section of the spherical stellar wind from the scattering region. That is,
the stellar wind is described by
$$
\v v_{disk}(\rho,\theta_y) = \cases{v_\infty (1-\rho^{-1})^\beta
\hat\v r &${\rm if}~\theta_o\le\theta_y\le\pi-\theta_o$  \cr
0&${\rm if~otherwise}$  \cr}
\eqno(4.3) $$
where $\theta_o$ is the opening angle of the disk wind.
Here, the normal direction of the disk is chosen to be the $y$-axis and the
$z$-axis coincides with the binary axis. The angle $\theta_y$ is defined as
the angle making with the $y$-axis, \ie,
$$\theta_y\equiv \tan^{-1}{\sqrt{x^2+z^2} \over y}\eqno(4.4)$$
The density law is given in a similar way as in Eq.~(4.3).

In Fig.~9 we display the numerical results from the disk wind with an
opening angle $\theta_0=\pi/4$. The results are displayed separately for the 
two cases. In the first case the observer sees the symbiotic system from the 
equatorial plane (Fig.~9a) and in the other case the observer is on the 
polar direction (Fig.~9b). Even though the azimuhtal symmetry about the binary axis
is broken, the dominant polarization turns out to be either along the binary axis
or perpendicular to it, and hence we keep the convention of the polarization
sign to denote the direction of polarization.

In both the cases the profiles and the degree of polarization
are qualitatively similar to those of the simple Gaussian case. The plausible
explanation for the results is that in the range where the single scattering
approximation is a valid one the same Doppler-$\tau_s$ contour
diagram (Fig.~2) applies and therefore a similar argument leads to the
qualitatively same results.

\subsec{4.4 Bipolar Flow}

As another application of non-spherical wind models, we investigate a
bipolar wind, which is simplified as a truncated spherical
wind model in a similar way to the disk wind case described in the
previous subsection. The wind symmetry axis is assumed to be perpendicular
to the binary axis. Therefore, the velocity law is
$$
\v v_{bipol}(\rho,\theta_y) = \cases{0 &${\rm if}~\theta_o\le\theta_y\le\pi-\theta_o$ \cr
v_\infty (1-\rho^{-1})^\beta \hat\v r &${\rm if~otherwise}$  \cr}
\eqno(4.5)$$

The numerical results are displayed in Fig.~10, where the wind opening 
angle $\theta_o=\pi/4$. In the left panel, the observer's line
of sight is perpendicular both to the bipolar wind direction and to
the binary axis, whereas in the right panel the observer lies in the
direction of the wind.

Unlike the disk wind case, we do not see any polarization flip in the
bipolar wind model. This is again explained in a similar way to the
case of the extended ionized region, where a large fraction of
the scattering region responsible for polarization flip in Fig.~2 is 
excluded in the bipolar wind model. Therefore, the overall behavior
of the polarization is characterized by a large degree of polarization
in the blue part of the feature and the existence of residual polarization
and flux in the red most part.

\subsec{4.5 Synthetic Profiles}

Some of symbiotic systems exhibit complex features in their emission lines.
The detailed physical nature of the emission line region of a symbiotic system
is controversial and the velocity scale associated with the stellar wind 
around the hot star is sometimes of order $10^3~\km~\s^{-1}$(\eg Vogel \&
Nussbaumer 1994). 
Osterbrock(1970) proposed that synthetic line profiles composed of
triangular shape may approximate the observed profiles quite well in the
planetary nebula IC 418.
We also use single-peaked and double-peaked triangular synthetic profiles
to compute the profiles and polarization of the Raman scattered lines.

The parameters necessary to describe the profiles are the widths and the
peak values. In this work, the half width at the bottom of the
single-peaked triangular profile is assumed to be $30~\km~\s^{-1}$. 
For the double-peaked triangular profile we superpose the two single-peaked
triangular profiles with different peak values. The ratio of the peak
values in the double-peaked profile is taken to be 0.3.

In the right panel of Fig.~11 are displayed the Monte Carlo results for
a single-peaked profile. 
The input profiles are shown in a small box at upper right corner
in each panel of Fig.~11.
As stated in Paper I synthetic profiles can
be decomposed into $\delta$-function profiles, which is the basis of the
Green function formalism.  In this regard,
we see a large degree of polarization of the blue part of the
feature, reminiscent of the simple Gaussian case discussed in section
4.1.

In the left panel of Fig.~11 we show the corresponding results for
a double-peaked profile. There are two peaks in the degree of polarization
corresponding to the blue wing of each triangular peak. Therefore, the
locations of the peaks in the degree of polarization shift to the
blue with respect to those of the scattered fluxes. The amount of the
shift depends mainly on the wind terminal velocity, and it increases
as the wind velocity gets larger. 
Hence, the wind terminal velocity can be constrained by the shift
of the peaks in the polarized flux with respect to the total scattered
flux.

\sec{5. Effect of Binary Motion}

It is expected that the hot component and the cool giant of a symbiotic
star revolve around their center of mass with a velocity
$$v_{rev}=\sqrt{G\mu/a}=30 \left({\mu_1\over a_1}\right)^{1/2}~\km~\s^{-1},
\eqno(4.6)$$
where $\mu_1=\mu/(1~M_\sun)$ is the reduced mass of the hot component and the
cool giant in terms of the solar mass and $a_1=a/(1~{\rm AU})$ is the
separation of the two components in the astronomical unit. This velocity
scale is comparable to the terminal wind velocity and therefore may affect
significantly on the profiles and the polarization of the Raman scattered
features. However, it is known phenomenologically that the typical separation 
$a$ of the `S' type symbiotic stars is very different from that of the `D' type
symbiotics (\eg Iben \& Tutukov 1996).
In this respect, the spectropolarimetry monitored for
an extended term will reveal a good deal of information about kinematics.
In this section,
we investigate the effect of the orbital motion of
the hot component with respect to the cool giant on the profiles and
the polarization of the Raman scattered features.

In Fig.~12 we draw the Doppler factor-$\tau_s$ contours analogous to
Fig.~2, where the hot component is moving in $-y$ direction with 
velocity $v_{rev}= 0.6\ v_\infty$, and the observer's line of sight
coincides with $x$-direction. 
The $\tau_s$ contour remains the same because 
it is assumed that the wind structure around the cool giant remains spherically
symmetric in the rest frame of the cool giant and the scattering
cross section is insensitive to the velocity scale of order $v_\infty=20
~\km~\s^{-1}$.
The diagram is plotted in the rest frame of the cool giant. Due to the
enhancement of the Doppler factor, the relative motion of the cool
giant with respect to the observer is not important compared with the
motion of the photon source relative to the scatterers.

Notable changes are seen in the structure of the Doppler factors. The contour
corresponding to zero Doppler factor is not spherical any more and
the sphere connecting the 
two stars may be decomposed into the upper hemisphere
with positive Doppler factors and the lower hemisphere with negative Doppler
factors. The important point to note about this sphere is that in the case of 
simple Gaussian case (see section 4.1) the polarization flip occurs 
around this sphere,
which corresponds to the central part of the scattered feature. Therefore,
we can expect from this diagram that there is possibility of polarization 
flip in the red part and/or the blue part of the scattered feature.

The numerical results corresponding to the situation depicted in Fig.~12 
are presented in Fig.~13, where the observer's line of sight
is perpendicular to the orbital plane. Here, we also present the position
angle because the symmetry about the binary axis is broken with the 
introduction of the relative motion of the hot component with respect to 
the cool giant. 
The position angle is not defined when the degree of polarization is 0.
Therefore in the extreme red part where there is so small
an amount of flux not to be assigned measurable polarization, the position
angle is highly uncertain and does not yield any quantitative information.

When the scattering optical depth $\tau_0=1$ and
$v_{rev}=0.6\ v_\infty$, the polarization flip occurs around the
red part. The degree of polarization is strong in the blue part of the 
feature. The position angle remains almost perpendicular to the binary 
axis in this part and parallel in the red part of 
the feature. The polarization component perpendicular to the binary axis
forms a double-peaked structure and therefore with the parallel component
the overall polarization shows a triple-peaked structure. The triple-peaked
structure is also seen in the polarized flux. 

In the total scattered flux no obvious multiple structures are seen in
these cases. The spectropolarimetry provided by Harries \& Howarth (1996)
shows that many symbiotic stars also have the multiple peak structures 
in the flux as well as in the polarization and the polarized flux. In the
simple Gaussian case with small optical depth ($\tau_0\le 0.5$) there
is a faint double peak structure in the flux. However, the small parameter
space is not enough to explain the much more complicated phenomena shown
by the observation. This
strongly implies that the simple Gaussian profile of the incident radiation 
is difficult to explain the complicated behavior of the spectropolarimetry.


It is also noted that the polarization flip occurs at a scattering optical
depth $\tau_0=1$. This is in contrast with the simple Gaussian case, where
the polarization flip at the center part of the feature is obvious only when
$\tau_0\go 5$.  This is particularly interesting considering
the spectropolarimetry data on the symbiotic systems including RR Tel, where
the polarization flip is clearly seen in the red part of the Raman-scattered
features (\eg Espey \etal 1995, Schmid \& Schild 1990, Schild \& Schmid 1996,
Harries \& Howarth 1996).

When the orbital velocity exceeds the wind terminal velocity, then
the polarization flip is also shown in the blue part, which is illustrated in
Fig.~14. Here, we adopt $v_{rev}=3\ v_\infty$. The polarization flip 
is seen in the red part and also in 
the blue part with small flux. This behavior may be explained using a similar
argument mentioned in the preceding paragraphs with Fig.~11. 

In Fig.~15, we show the numerical results when the observer's
line of sight lies on the orbital plane. The results are similar to those of
the simple Gaussian case, and therefore no polarization flip around the 
red part occurs in this case. This dramatically different behavior can
be also explained using the Doppler factor-$\tau_s$ diagram. 
When the observer's line of sight lies on the orbital plane, the section 
of the Doppler factor-$\tau_s$ diagram including the
binary axis and perpendicular to the observer's line of sight 
coincides with Fig.~2, and the polarization behavior
is not significantly different from that of the simple Gaussian case.

\sec{6. Observational Implications}

Observational and evolutionary properties of symbiotic stars are 
extensively summarized by Iben \& Tutukov (1996), according to whom
symbiotic stars are rather inhomogeneous group showing a large range of
variabilities and physical conditions. Detailed analysis of
the UV emission lines of symbiotic stars also implies that kinematics
around the emission line region may not be described by a simple unique
model. 

In the optically thin limit, the ratio of the $\lambda 1032$ flux to that of
the $\lambda 1038$ flux in the O~VI doublet is 2, which is the ratio of the oscillator
strengths of the two transitions. However, this ratio is altered when the
medium gets optically thick, where the stronger lines have more difficulty
escaping the region. Far UV observations show that for the doublets
including C~IV $\lambda\lambda~ 1548,~ 1551$, N~V $\lambda\lambda~1238,~
1241$ the ratios are usually less than 2, and sometimes less than 1 
(\eg Vogel \& Nussbaumer 1994), which confirms the complicated nature of
the emission line region.

The Raman scattered features provide a special tool for 
investigation of the physical conditions of a symbiotic star through
spectropolarimetry. Schmid \& Schild (1994) provided spectropolarimetric
data of some symbiotic systems, and more recently Harries \& Howarth (1996)
presented more data with enhanced resolution. The polarization behavior
is also very heterogeneous and has complex structures. Schmid \& Schild (1994)
divided them into the three types and described the basic points.
The extended ionized region may have definite effects on the polarization 
behavior displayed in the observational data  as shown by Schmid (1996).

It has been pointed out by several researchers including Schmid (1996),
and Harries \& Howarth (1996,1997) that the position angle varies
throughout the Raman scattered features. This implies that the most
general model should include those producing the polarization direction
other than the binary axis. Because the polarization behavior based
on the models possessing a symmetry with respect to the binary axis 
does not show any rotation of the position angle, it has been proposed 
to consider non-spherical models. 

Harries \& Howarth (1996) pointed out that most of the Raman scattered 
features show multiple peak structures in the flux and in the polarized flux,
and in particular double and triple peaks are found very frequently. 
These structures are maybe coupled with the complicated profiles of the
UV emission lines including a P-Cygni type profile. If the incident
UV line profile has a double-peaked structure, the similar structure is
obtained in the scattered profile and in the polarized flux. However,
the location of the peaks may differ depending sensitively on the
wind terminal velocity. In this regard, it is
necessary to consider various input profiles for reproducing the 
characteristics of the observed Raman scattered fluxes.

Some fraction of symbiotic stars show polarization flip around the red wing
of the scattered features, which is produced in the Monte Carlo computation
when the orbital motion of the hot component about the cool giant is
included. Typical parameters of symbiotic stars imply that the rotation
velocity is almost comparable to the wind terminal velocity (\eg Eq.~(4.6)),
and the polarization flip is shown clearly when the orbital plane coincides
with the sky plane. The orbital motion can be observed by a careful
monitoring of the position angle variation (\eg Schild \& Schmid 1996). 
However, we predict that the polarization flip will be seen irrelevant 
of the orbital phase, 
because the inclination angle of the orbital plane with respect to the line of
sight does not change with the orbital motion, 

Furthermore, the polarization flip is shown in the blue part when the orbital
velocity exceeds the wind terminal velocity. In this case, the polarized
flux shows a clear triple-peaked structure. 
Therefore, an independent observational determination
of the orbital velocity will provide a good constraint to the wind
terminal velocity, which is proportional to the mass loss rate of the cool
giant.

It is proposed that the effect of the binary motion
may leave an important signature in the polarization and polarized
flux. It will be also interesting to investigate the combined effects
of the binary motion and the non-Gaussian input profiles with different
ionization structures. However, the introduction of the binary motion
breaks the azimuthal symmetry and requires a good amount of computation
time and a huge volume of parameter space to be examined. Therefore, 
independent determination
of the kinematic and the dynamic parameters by the UV observations 
and monitoring of 
spectropolarimetry are expected to unveil the physical nature of 
symbiotic stars.

\sec{7. Summary}

The Raman scattered features are quite unique to symbiotic stars so far 
and can be used as a powerful diagnostic of the physical conditions including
the ionization structures and the kinematics on the orbital motion and the
stellar wind structures. They possess several distinguished characteristics.
The first and the most important point is that they are usually strongly
polarized and therefore spectropolarimetry reveals a significant
amount of information
about the kinematics and the scattering geometry. Related with this
is that they are composed of purely scattered photons without dilution from the
direct flux retaining the vivid information
on the scattering geometry. The third point is that
the inelastic nature of the Raman process provides the enhancement of the
profile width by a factor of $\lambda_f /\lambda_i$, which amounts to
almost an order of magnitude in the case of the O~VI doublets.

In this study, we reproduce the general properties of the Raman scattered
emission lines, which are found in the works of many researchers
including Schmid (1995, 1996) and Harries \& Howarth (1997). The
spherical wind model usually gives a large degree of polarization in
the blue wing and the polarization flip around the center of the
feature depending on the scattering optical depth. 

If we take the orbital motion of the hot component around the cool 
giant into consideration, then the polarization flip is seen to shift to the 
red wing when the orbital plane is near to the sky plane. This
effect is conspicuously seen when the orbital velocity is almost 
comparable to the terminal velocity of the stellar wind, giving insight
into the spectropolarimetric data of several symbiotic stars including
RR Tel (\eg Espey \etal 1995, Harries \& Howarth 1996, Schmid 1996). 
When the orbital velocity exceeds greatly the terminal velocity, the 
polarization flip is expected to occur both in the blue wing and in 
the red wing. We also expect that various input profiles proposed
by generic dynamical models can be combined to produce the complex
features in the Raman scattered lines.

Much parameter space remains to be investigated by
numerical methods, including the parameters $\dot M$, $\beta$, $X_H$, 
$v_\infty$, $v_{rev}$ to list a few, and it is hoped that the numerical
calculation may provide a useful constraint to existing dynamical models. 
It is generally concluded that spectropolarimetry
will remain a useful tool to delve into the symbiotic system.

\sec{Acknowledgements}

We are grateful to Profs. M. G. Lee, S. S. Hong  for helpful discussions. 
We are also grateful for the valuable comments by the referee, Dr. H. Schmid.
H. W. L gratefully acknowledges the support from the
Research Institute for Basic Sciences at the Seoul National University.
K. W. L was supported in part
by KOSEF directed research grant 94-0702-04-01-3 and KOSEF
non-directed research grant 971-0203-013-2.

\newdimen\refindent
\refindent=0.5truecm
\def\ref#1#2#3#4#5{
 \let\qzero=0
 \let\qpoint=.
 \let\qpaper=1
 \def\qthree{#3}
 \def\qfour{#4}
 \def\qfive{#5}
 \filbreak\hangindent\refindent\hangafter=1\noindent
 \if\qpoint\qfour
  {#1, 19#2, #3}
 \else
  \if\qzero\qthree
   {#1, 19#2, In preparation}
  \else
   \if\qzero\qfour
    \if\qzero\qfive
     {#1, 19#2, #3, in press}
    \else
     {#1, 19#2, #3, p.#5}
    \fi
   \else
    \if\qzero\qfive
     {#1, 19#2, {#3}, in press}
    \else
     {#1, 19#2, {#3}, {#4}, #5}
    \fi
   \fi
  \fi
 \fi
}
\def\aap{A\&A}
\def\aas{A\&AS}

\def\apj{ApJ}

\def\apjs{ApJS}

\def\mn{MNRAS}

\def\jkass{Journal of Korean Astronomical Society Supplement}
 



\vfill\eject
\sec{REFERENCES}
\ref{Allen D. A.}{80}{\mn}{190}{75}
\ref{Berestetskii V.B., Lifshitz E.M., Pitaevskii L.P.}
{71}{{Relativistic Quantum Mechanics.} Pergamon Press,
New York}{.}{}
\ref{Espey B. R. \etal}{95}{\apj}{454}{L61}
\ref{Harries T.J., Howarth I.D.}{96}{\aas}{119}{61}
\ref{Harries T.J., Howarth I.D.}{97}{\aas}{121}{15}
\ref{Iben Jr. I., Tutukov A. V.}{96}{\apjs}{105}{145}
\ref{Isliker H., Nussbaumer H. Vogel M.}{89}{\aap}{219}{271}
\ref{Lee H. W., Blandford R. D. Western R.}{94}{\mn}{267}{303}
\ref{Lee H. W., Lee K. W.}{96}{\jkass}{29}{S249}
\ref{Lee H. W., Lee K. W.}{97}{\mn}{287}{211 (Paper I)}
\ref{Mueller B. E. A., Nussbaumer H.}{85}{\aap}{145}{144}
\ref{Nussbaumer H., Vogel M.}{87}{\aap}{182}{51}
\ref{Osterbrock D. E.}{70}{\apj}{159}{823}
\ref{Pereira C. B., Vogel M., Nussbaumer H.}{95}{\aap}{293}{783}
\ref{Schmid H. M.}{89}{\aap}{211}{L31}
\ref{Schmid H. M.}{92}{\aap}{254}{224}
\ref{Schmid H. M.}{95}{\mn}{275}{227}
\ref{Schmid H. M.}{96}{\mn}{282}{511}
\ref{Schmid H. M., Schild H.}{90}{\aap}{236}{L13}
\ref{Schmid H. M., Schild H.}{94}{\aap}{281}{145}
\ref{Schild H., Schmid H. M.}{96}{\aap}{310}{211}
\ref{Seaquist E. R., Taylor A. R., Button S.}{84}{\apj}{284}{202}
\ref{Vogel M., Nussbaumer H.}{94}{\aap}{284}{145}
\vfill\eject
\sec{FIGURE CAPTION}

Figure~1.- A  schematic geometry of a symbiotic star system for the
Monte Carlo calculation.
The $z$-axis is chosen to be the binary axis and the observer's line
of sight coincides with $x$-axis.
The hot star is represented by the circle on
the right side, and the bigger circle in the origin is the cool
giant. A spherically symmetric stellar wind is shown by radial (small) 
arrows. The vectors $\vec\rho_i$ and $\vec\rho_{i+1}$ represent
the $i$-th and the $i+1$-th scattering sites respectively and $s(\tau_i)$
is the spatial distance between the two scattering sites. $b$ is the
impact parameter of the photon trajectory with respect to the cool giant.
\vskip 10pt
Figure~2.- A contour map of the Doppler factors and the scattering
optical depths of incident photons from the hot star. The stellar
wind is a spherical one defined in section~2.2. The Doppler factor
is defined by $DF \equiv \hat \v k_i \cdot \v v /c$ and the total
scattering optical depth $\tau_s\equiv \int ds~n(s)(\sigma_{Ray}+\sigma_{Ram})$
. The adopted parameters are described in Paper I.
\vskip 10pt
Figure~3.- Comparison of the single scattering approximation with
corresponding Monte Carlo calculation. The dotted lines represent the
Monte Carlo result and the result from the single scattering 
approximation is shown by the solid lines.
\vskip 10pt
Figure~4.- The Rayleigh and Raman reflected fluxes from a finite
slab of total scattering optical depth $\tau_s=0.5,1,5,10$ as a
function of the number of scattering. The light lines represent
the result for the case $R_{Ray}=0.2$ and the case for $R_{ray}=0.8$
is shown by the thick lines. The solid lines are for the Raman
scattered flux and the dotted lines are for the Rayleigh case.
\vskip 10pt
Figure~5.- The Rayleigh and the Raman reflected fluxes from a slab of 
the total scattering optical depth $\tau_s=10,~R_{Ray}=0.8$ as in
Fig.~4(d) with semi-quantitatively described $f_{Ram}(n),~f_{Ray}(n)$
in section 3.2. The solid lines are for the Monte Carlo results, and
the dotted lines give $f_{Ram}(n),~f_{Ray}(n)$ computed using Eqs.
(3.3)-(3.6). The upper set of curves are Raman scattered lines and the
lower set corresponds to the Rayleigh flux.
In the bottom panel, $f_{Ram}(n)$, and $f_{Ray}(n)$ 
calculated from a different escape probability $\beta'(n)=
\exp(-n^{0.4})/2$ (Eq.~(3.7)).
\vskip 10pt
Figure~6.- The line profiles and the polarization for the simple
Gaussian case, where the UV emission line profiles are described by
a Gaussian with $T_{em}=10^4~\K$. See the text for more detail.
\vskip 10pt
Figure~7.- A typical ionization structure around the hot star of a 
symbiotic system, reproduced using the recipes provided by Seaquist
\etal(1984). The parameter $X_H$ defined  in Eq.~4.2 determines the
overall structure of the ionized region. Solid lines divide H~I region
for H~II. Dotted lines are obtained using a simpler equation for
the ionization front.
\vskip 10pt
Figure~8.- Monte Carlo results corresponding to the ionization
structures described in Fig.~7.
\vskip 10pt
Figure~9.- Monte Carlo results for the disk-wind models. \par
(a) The observer's line of sight lies in the equatorial plane of the
disk wind and the typical scattering optical depths $\tau_0=0.5,~1.0,~
10$. \par
(b) The observer's line of sight is perpendicular to the equatorial
plane.
\vskip 10pt
Figure~10.- Monte Carlo results for the bipolar wind models. \par
(a) The observer's line of sight is perpendicular both to the
binary axis and to the wind symmetry axis and the scattering 
optical depths $\tau_0=0.5,~1,~10$. \par
(b) The observer's line of sight coincides with the wind symmetry axis.
\vskip 10pt
Figure~11.- Monte Carlo results for the cases where the input
profiles are gives by single-peaked and double-peaked profiles.
The input profiles are shown in a small box at upper right corner
in each panel. \par
(a) Single-peaked profile with $\tau_0=1$ \par
(b) Single-peaked profile with $\tau_0=10$ \par
(c) Double-peaked profile with $\tau_0=1$ \par
(b) Double-peaked profile with $\tau_0=10$ \par
\vskip 10pt
Figure~12.- Doppler-$\tau_s$ contours analogous to Fig.~2 
including the relative motion
of the hot component with respect to the cool giant. The hot star
is assumed to move in $-y$-direction with a velocity $v_{rev}=0.6\
v_{\infty}$, where $v_\infty$ is the terminal velocity of the spherical
stellar wind around the cool giant.
\vskip 10pt
Figure~13.- Monte Carlo results corresponding to the case described
in Fig.~12. The observer's line of sight is perpendicular to the orbital
plane. \par
(a) The typical scattering optical depth $\tau_0=1$ \par
(b) The typical scattering optical depth $\tau_0=5$ \par
\vskip 10pt
Figure~14.- Monte Carlo results for with $v_{rev}=3\ v_{\infty}$. The wind
structure around the cool giant and the observer's line of sight are the same
as in Fig.~12. \par
(a) $\tau_0=1$, (b) $\tau_0=5$ \par
\vskip 10pt
Figure~15.- Monte Carlo results for the same geometry with Fig.~12,
where the observer's line of sight lies on the orbital plane. \par
(a) $\tau_0=1$, (b) $\tau_0=5$ \par
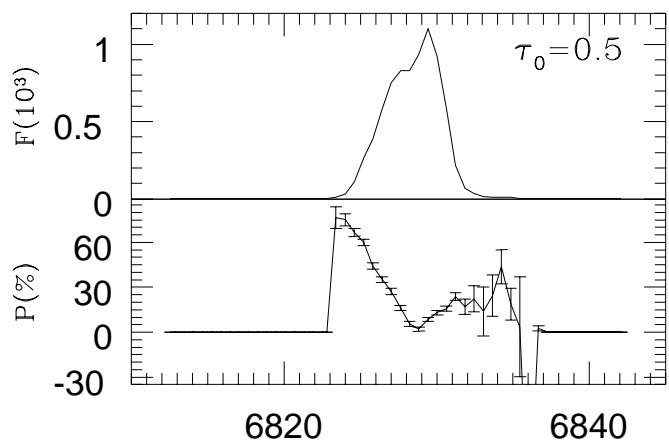
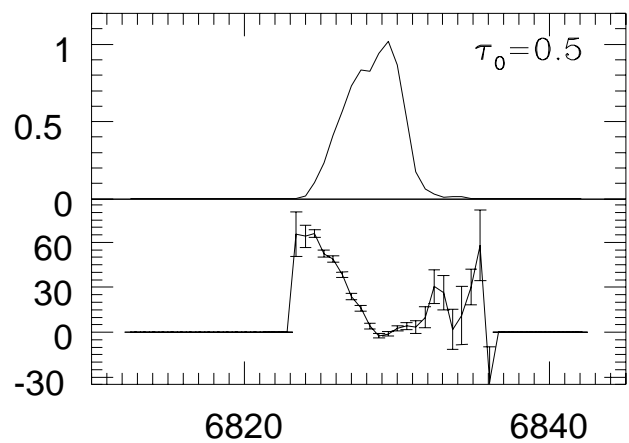
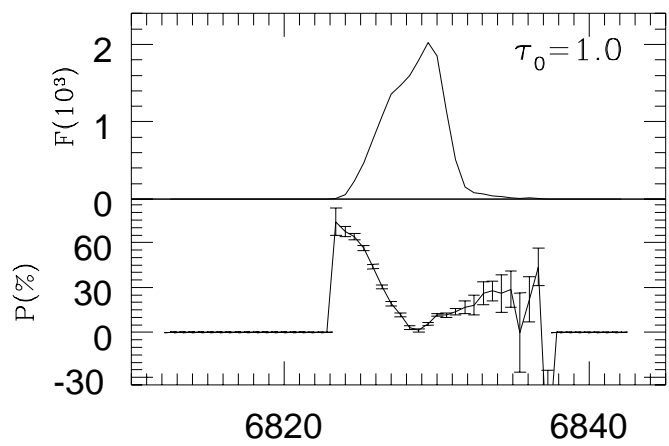
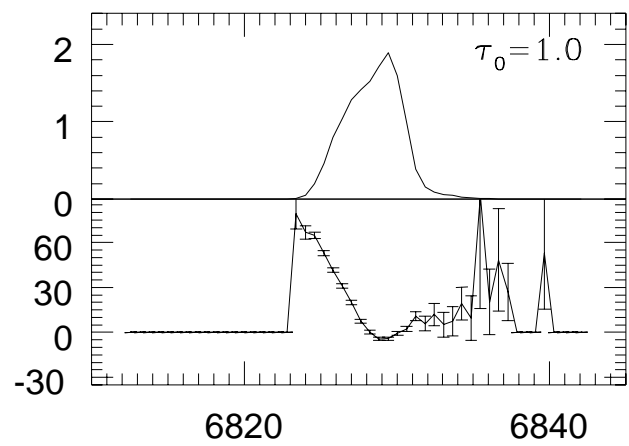
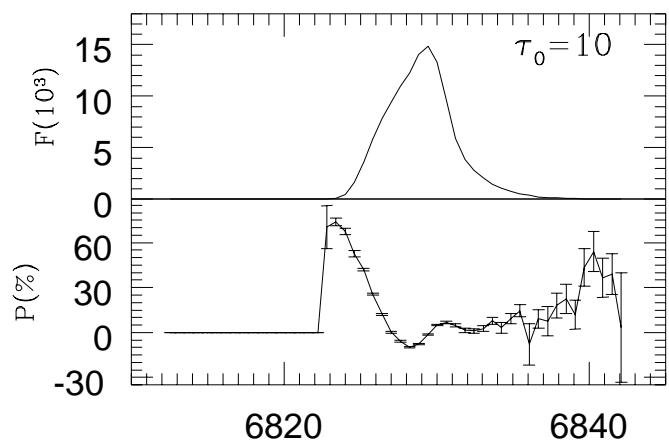
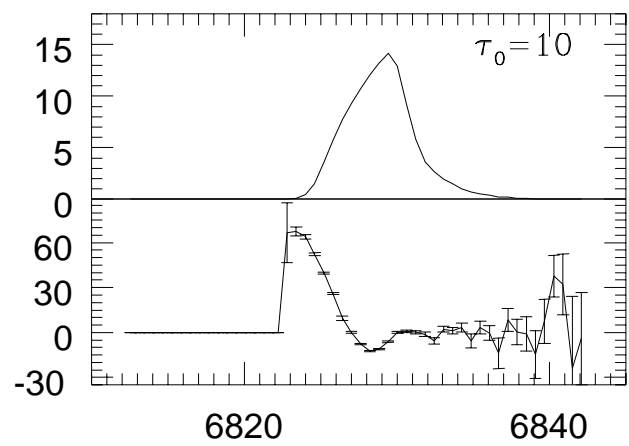
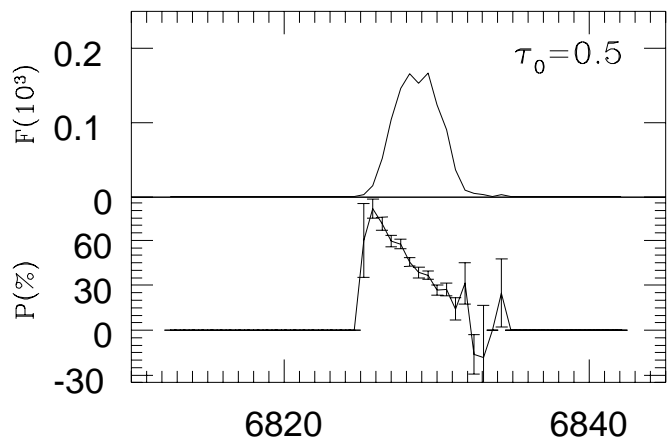
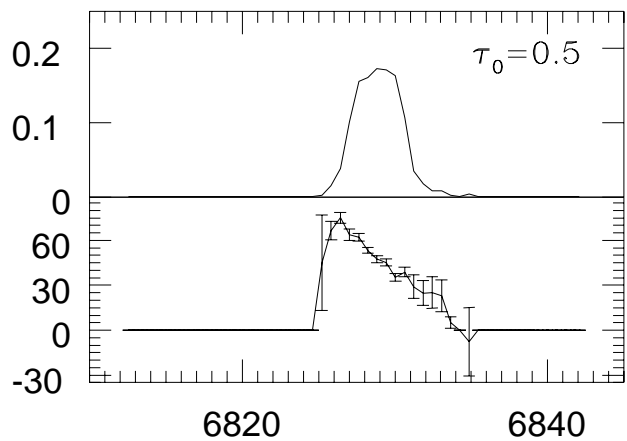
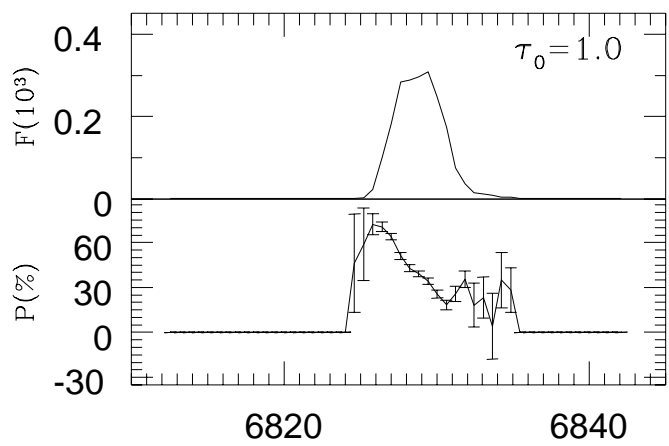
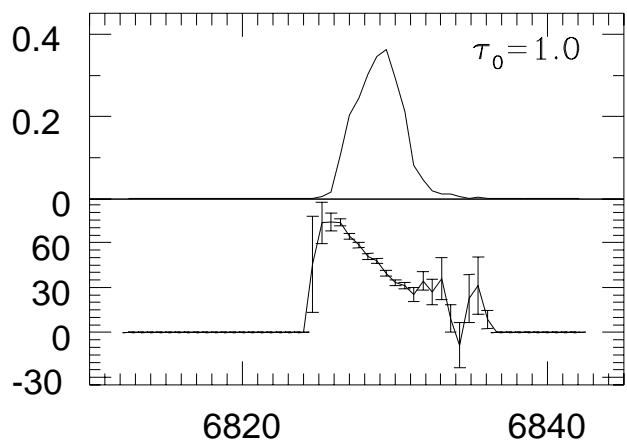
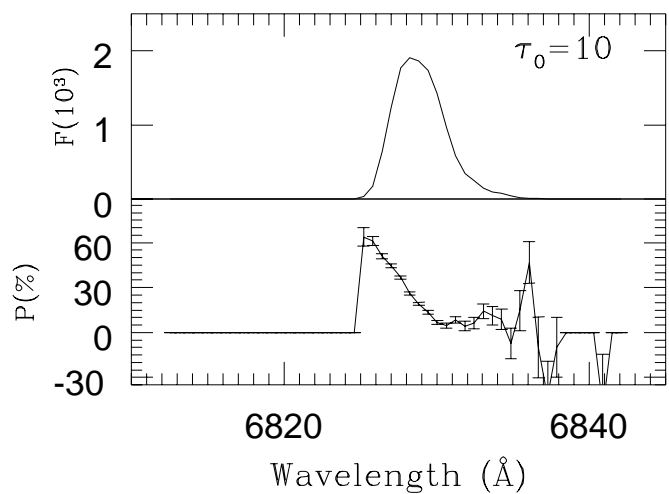
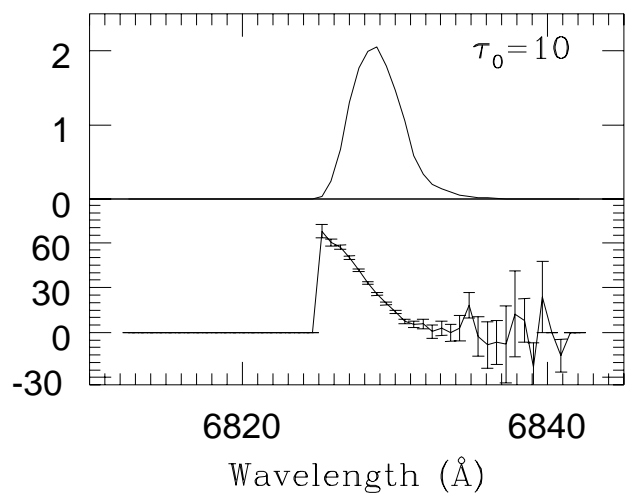
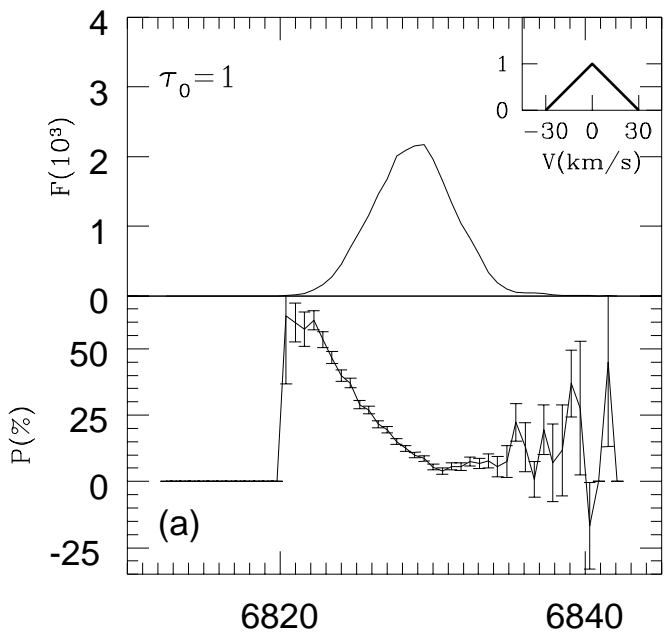
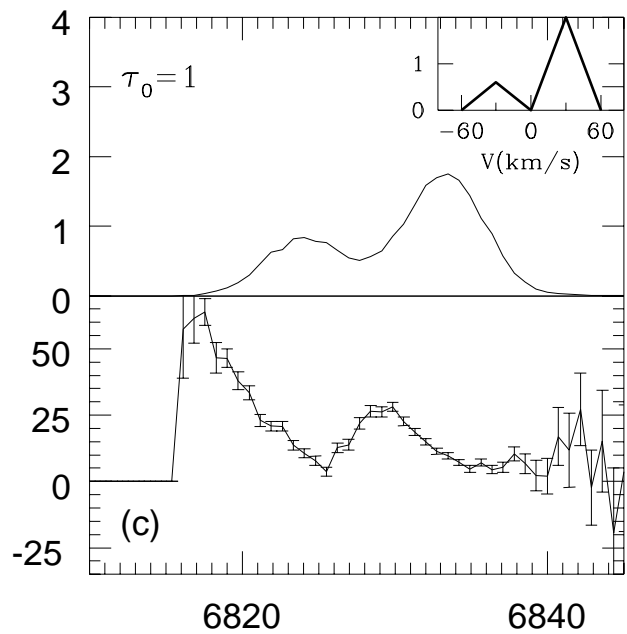
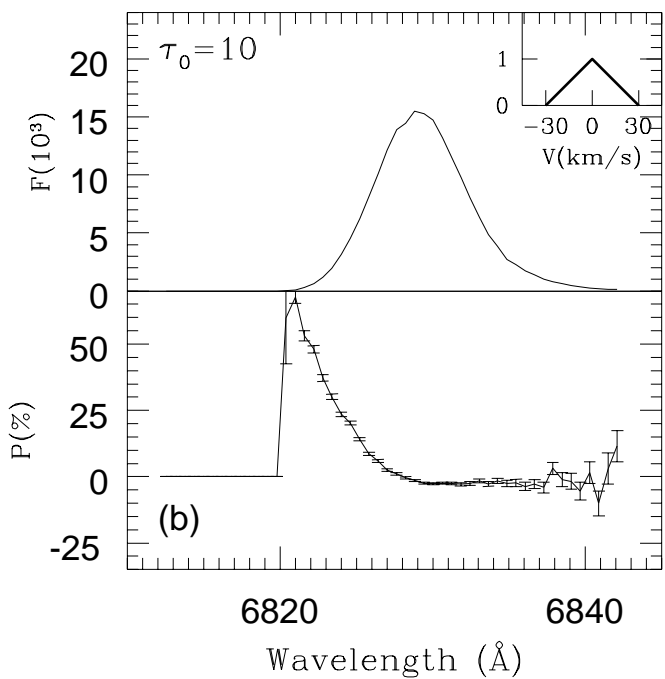
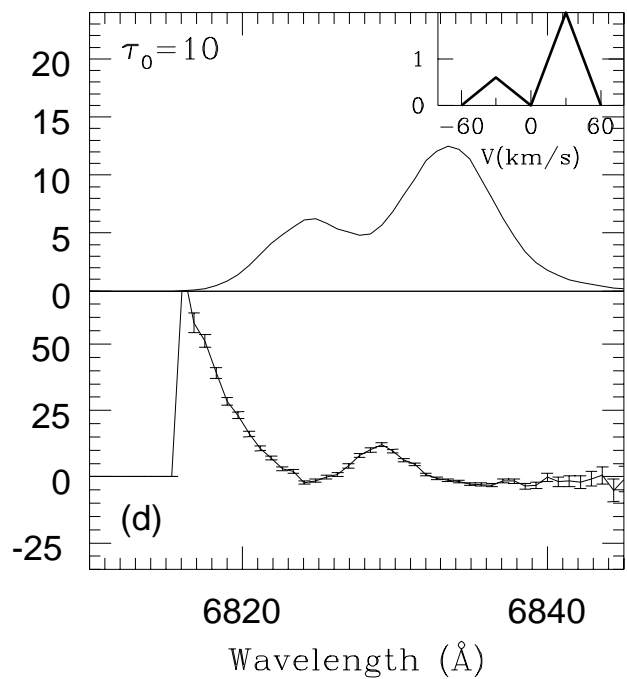
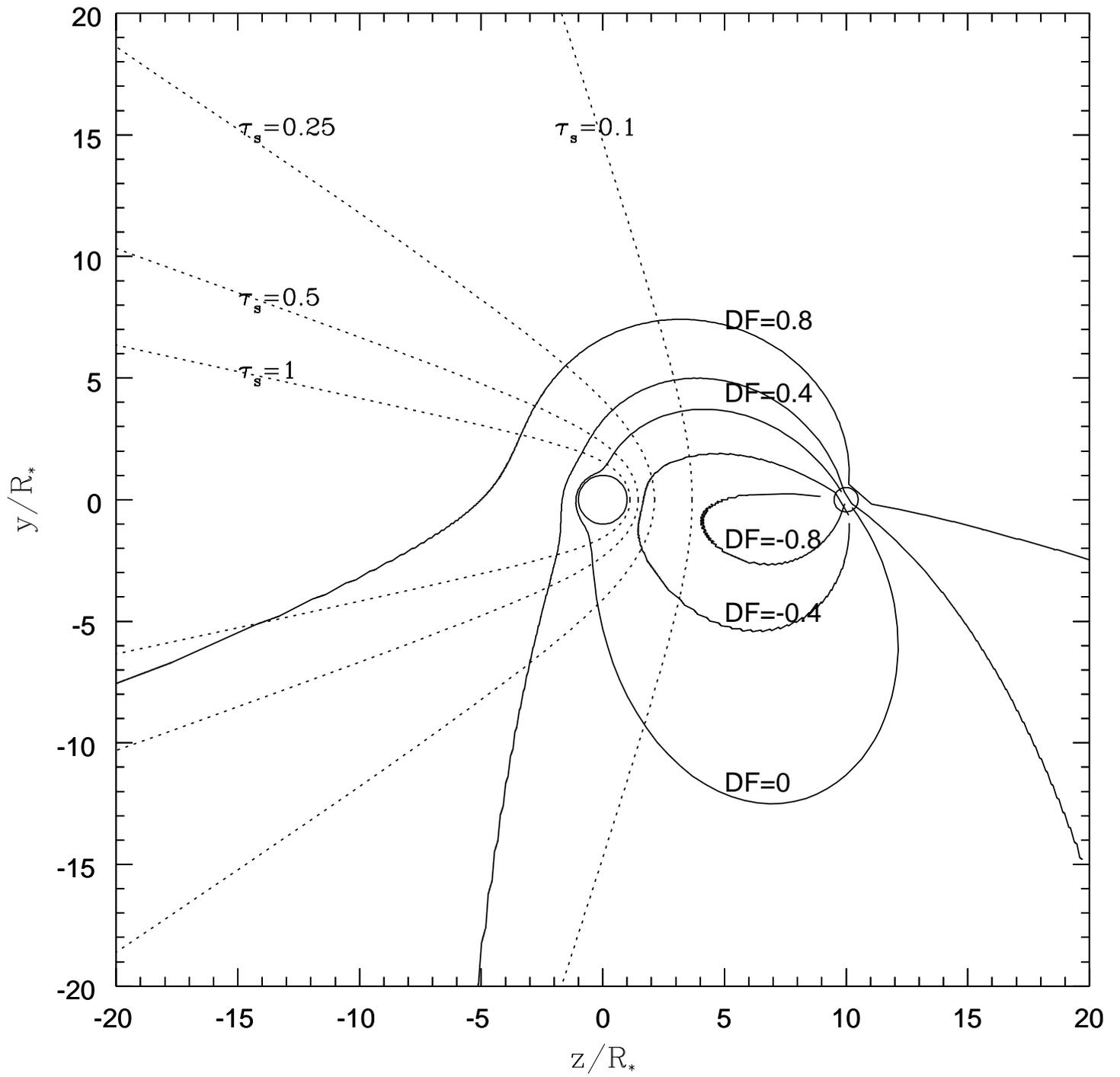
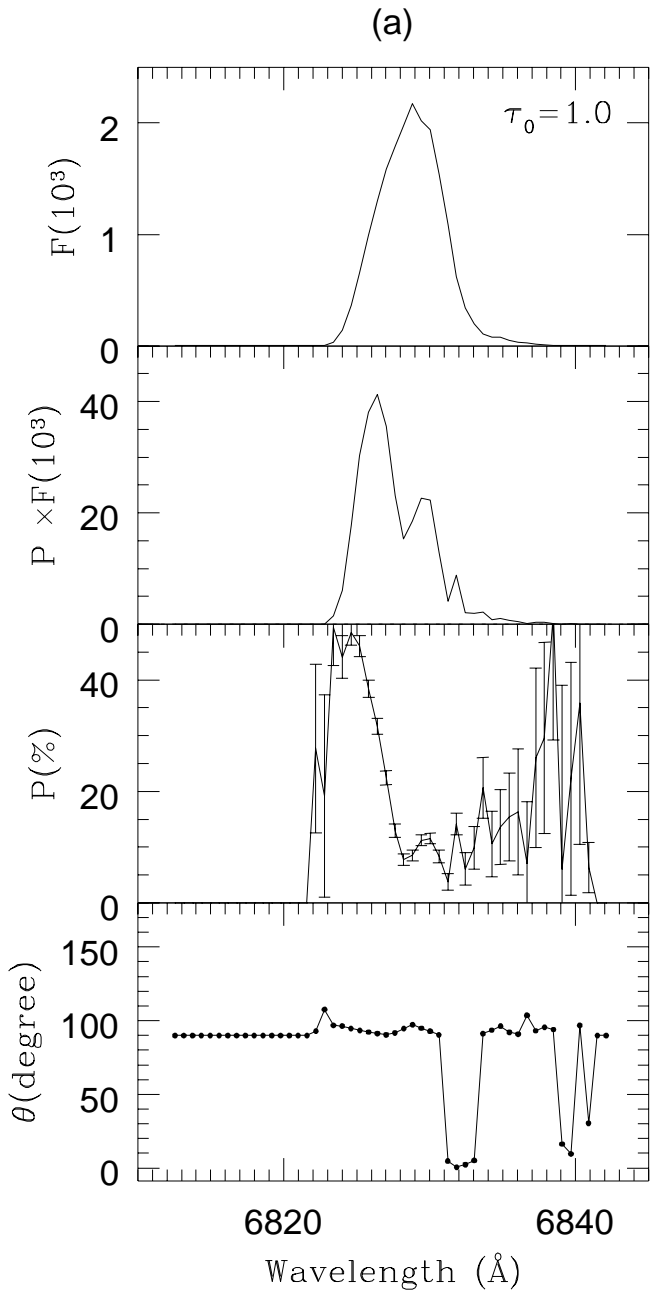
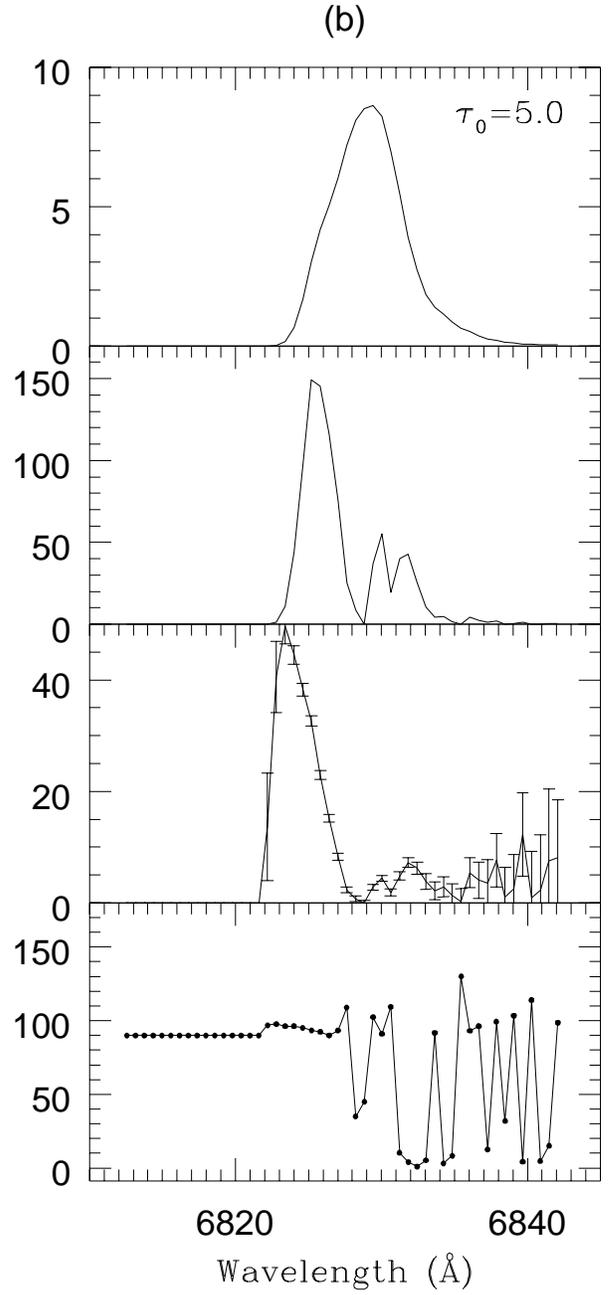
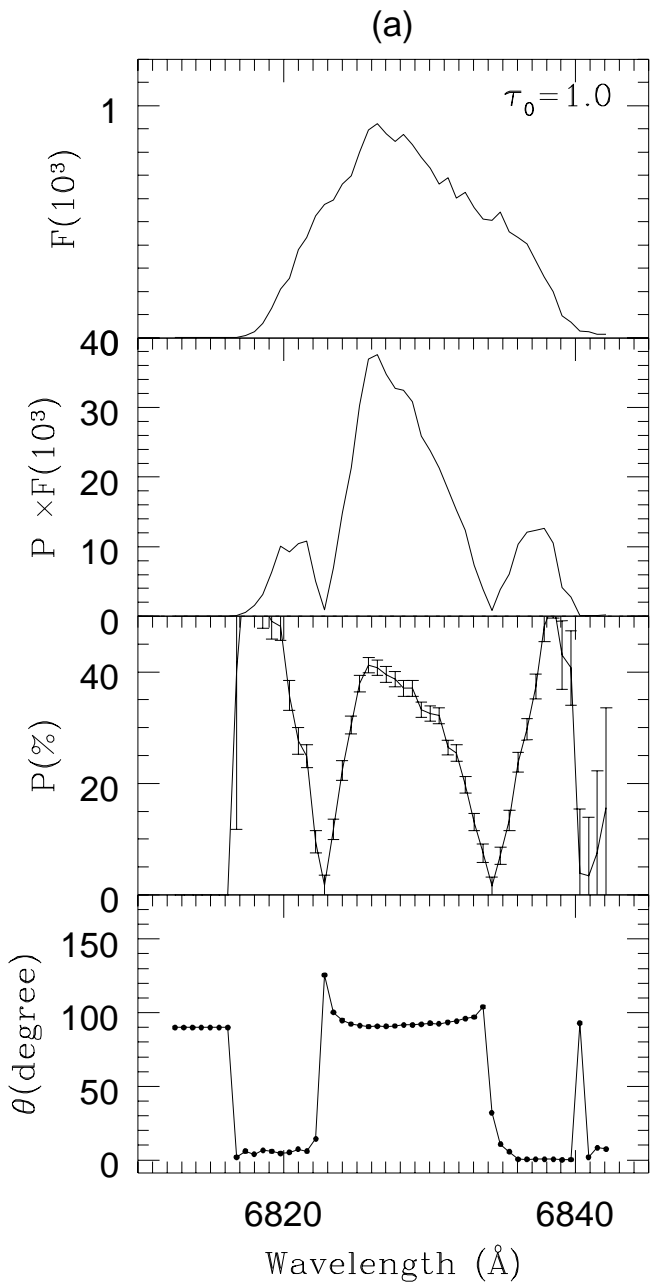
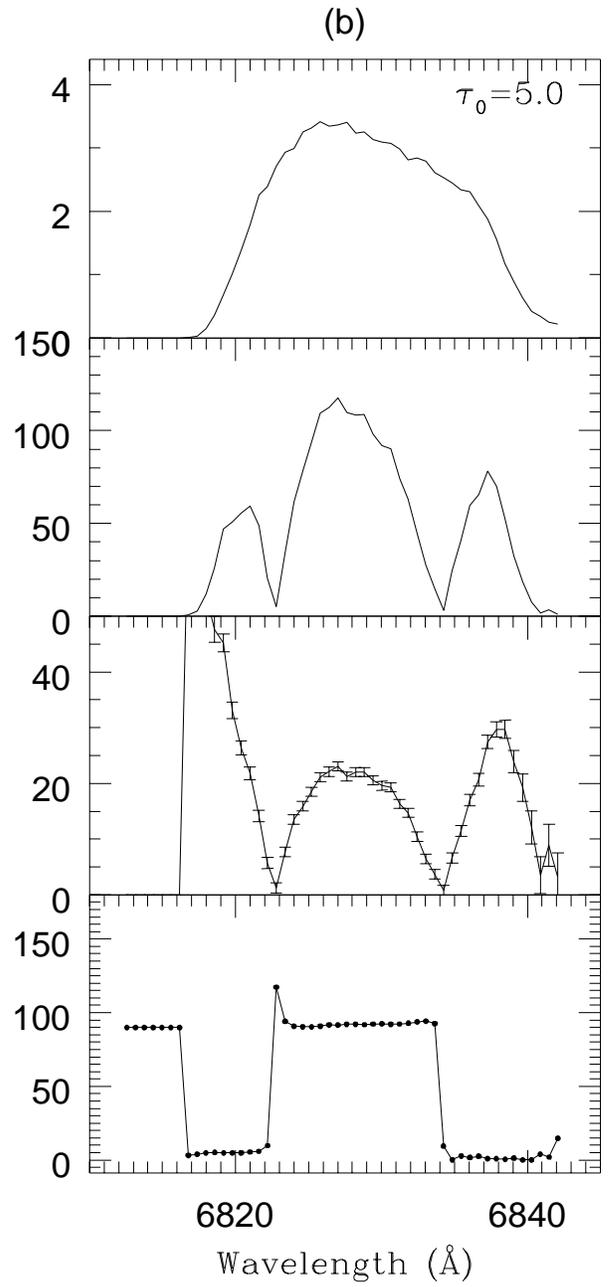
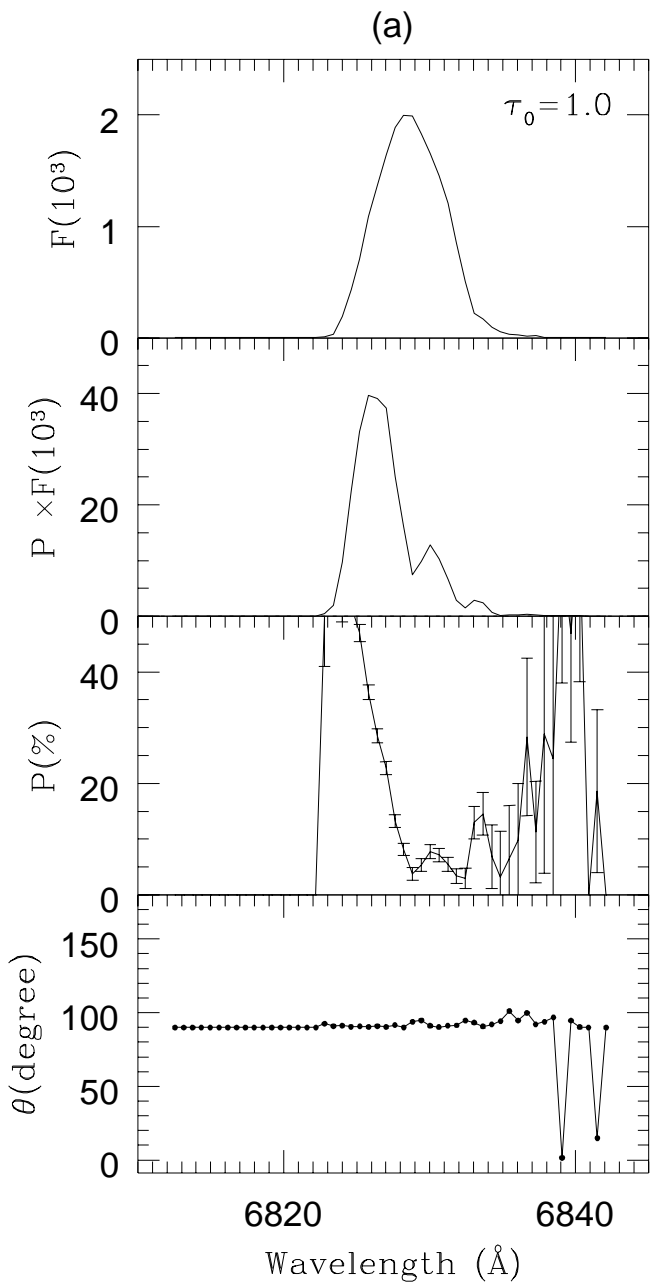
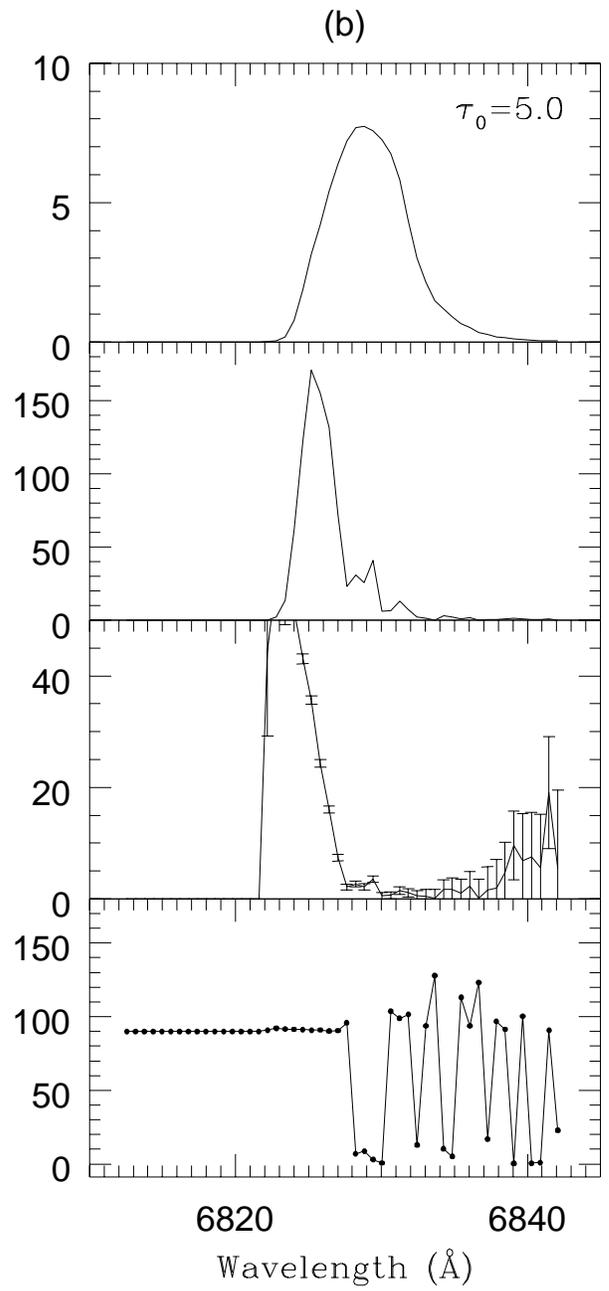
\end